\documentclass[12pt,aps,prb,preprint]{revtex4}

\usepackage{amsmath}
\usepackage{amssymb}
\usepackage{graphicx}
\usepackage{bm}

\draft

\begin{document}

\title{Numerical simulation of the double slit interference with ultracold
atoms}

\author{Michel Gondran}
\email{michel.gondran@chello.fr} \affiliation{EDF, Research and
Development, 1 av.\ du General de Gaulle, 92140 Clamart, France}

\author{Alexandre Gondran}
\email{alexandre.gondran@utbm.fr} \affiliation{University of
Technology Belfort-Montbéliard, 90010 Belfort cedex, France}

\begin{abstract}
We present a numerical simulation of the double slit interference
experiment realized by F.\ Shimizu, K.\ Shimizu and H.\ Takuma
with ultracold atoms. We show how the Feynman path integral method
enables the calculation of the time-dependent wave function.
Because the evolution of the probability density of the wave
packet just after it exits the slits raises the issue of the
interpreting the wave/particle dualism, we also simulate
trajectories in the de Broglie-Bohm interpretation.
\end{abstract}

\maketitle

\section{Introduction}

In 1802, Thomas Young (1773--1829), after observing fringes inside
the shadow of playing cards illuminated by the sun, proposed his
well-known experiment that clearly shows the wave nature of
light.\cite{Young_1802} He used his new wave theory to explain the
colours of thin films (such as soap bubbles), and, relating colour
to wavelength, he calculated the approximate wavelengths of the
seven colours recognized by Newton. Young's double slit experiment
is frequently discussed in textbooks on quantum
mechanics.\cite{FeynmanCours}

Two-slit interference experiments have since been realized with
massive objects, such as
electrons,\cite{Davisson,Jonsson,Merli,Tonomura_1989}
neutrons,\cite{Halbon,Rauch} cold neutrons,\cite{Zeilinger_1988}
atoms,\cite{Estermann} and more recently, with coherent ensembles
of ultra-cold atoms,\cite{Shimizu,Anderson} and even with
mesoscopic single quantum objects such as C$_{60}$ and
C$_{70}$.\cite{Arndt,Nairz}

This paper discusses a numerical simulation of an experiment with
ultracold atoms realized in 1992 by F.\ Shimizu, K.\ Shimizu, and
H.\ Takuma.\cite{Shimizu} The first step of this atomic
interference experiment consisted in immobilizing and cooling a
set of neon atoms, mass $m=3.349\,\times\,10^{-26}$\,kg, inside a
magneto-optic trap. This trap confines a set of atoms in a
specific quantum state in a space of $\simeq1\,$mm, using cooling
lasers and a non-homogeneous magnetic field. The initial velocity
of the neon atoms, determined by the temperature of the
magneto-optic trap (approximately~ $T=2.5~mK$) obeys to a Gaussian
law with an average value equal to zero and a standard deviation
$\sigma_v=\sqrt{\frac{k_BT}{m}}\simeq 1~m/s$; $k_B$ is Boltzman's
constant.

\begin{figure}[h!]
\begin{center}
\includegraphics[width=0.9\linewidth]{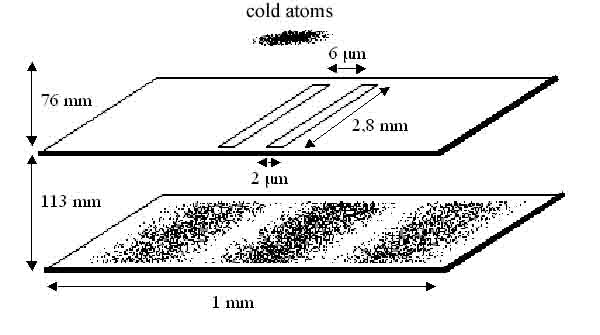}
\caption{Schematic configuration of the
experiment.}\label{fig:experience}
\end{center}
\end{figure}

To free some atoms from the trap, they were excited with another
laser with a waist of 30\,$\mu$m. Then, an atomic source whose
diameter is about $3 \times 10^{-5}$\,m in and $10^{-3}$\,m in the
$z$ direction was extracted from the magneto-optic trap. A subset
of these free neon atoms start to fall, pass through a double slit
placed at $l_1=76$\,mm below the trap, and strike a detection
plate at $l_2=113$\,mm. Each slit is $b=2\,\mu$m wide, and the
distance between slits, center to center, is $d=6\,\mu$m. In what
follows, we will call ``before the slits'' the space between the
source and the slits, and ``after the slits'' the space on the
other side of the slits. The sum of the atomic impacts on the
detection plate creates the interference pattern shown in
Fig.~\ref{fig:experience}.

The first calculation of the wave function double slit experiment
using electrons\cite{Jonsson} was done using the Feynman path
integral method.\cite{Philippidis}  However, this calculation has
some limitations. It covered only phenomena after the exit from
the slits, and did not consider realistic slits. The slits, which
could be well represented by a function $G(y)$ with $G(y)=1$ for
$-\beta\leq y \leq \beta$ and $G(y)=0$ for $|y|>\beta$, were
modeled by a Gaussian function $G(y)=e^{-y^2/2 \beta^2}$.
Interference was found, but the calculation could not account for
diffraction at the edge of the slits. Another simulation with
photons, with the same approximations, was done
recently.\cite{Ghose} Recently, some interesting simulations of
the experiments on single and double slit diffraction of
neutrons\cite{Zeilinger_1988} were done.\cite{Sanz_2002}

The simulations discussed here cover the entire experiment,
beginning with a single source of atoms, and treat the slit
realistically, also considering the initial dispersion of the
velocity. We will use the Feynman path integral method to
calculate the time-dependent wave function. The calculation and
the results of the simulation are presented in Sec.~II. The
evolution of the probability density of the wave packet just after
the slits raises the question of the interpretation of the
wave/particle dualism. For this reason, it is interesting to
simulate the trajectories in the de Broglie\cite{deBroglie} and
Bohm\cite{Bohm} formalism, which give a natural explanation of
particle impacts. These trajectories are discussed in Sec.~III.

\section{Calculation of the wave function with Feynman path integral}

In the simulation we assume that the wave function of each source
atom is Gaussian in $x$ and $y$ (the horizontal variables
perpendicular and parallel to the slits) with a standard deviation
$\sigma_0 = \sigma_x = \sigma_y = 10\,\mu$m. We also assume that
the wave function is Gaussian in $z$ (the vertical variable) with
zero average and a standard deviation $\sigma_z\simeq 0.3$\,mm.
The origin $(x=0,y=0,z=0)$ is at the center of the atomic source
and the center of the Gaussian.

The small amount of vertical atomic dispersion compared to typical
vertical distances, $\sim 100$ and 200\,mm, allows us to make a
few approximations. Each source atom has an initial velocity
$\textbf{v}=(v_{0x},v_{0y},v_{0z})$ and wave vector
$\textbf{k}=(k_{0x},k_{0y},k_{0z})$ defined as
$\textbf{k}=m\textbf{v}/\hbar$. We choose a wave number at random
according to a Gaussian distribution with zero average and a
standard deviation $\sigma_k =\sigma_{k_x}
=\sigma_{k_y}=\sigma_{k_z}=m \sigma_v /\sqrt{3}  \hbar \simeq 2
\times 10^8$\,m$^{-1}$, corresponding to the horizontal and
vertical dispersion of the atoms' velocity inside the cloud
(trap). For each atom with initial wave vector \textbf{k}, the
wave function at time $t=0$ is
\begin{eqnarray}
\psi_0(x,y,z;k_{0x},k_{0y},k_{0z})&= &\psi_{0_x}(x;k_{0x})
\psi_{0_y}(z;k_{0y}) \psi_{0_z}(z;k_{0z})
\nonumber \label{eq:psi0} \\
&=&(2\pi\sigma_0^2)^{-1/4}e^{-x^2/4\sigma_0^2}\,e^{ik_{0x}x}
\nonumber \\ {}&&{}\times
(2\pi\sigma_0^2)^{-1/4}\,e^{-y^2/4\sigma_0^2}\,e^{ik_{0y}y}
\nonumber \\
{}&&{}\times(2\pi\sigma_z^2)^{-1/4}\,e^{-z^2/4\sigma_z^2}
e^{ik_{0z}z}.
\end{eqnarray}

The calculation of the solutions to the Schr\"odinger equation
were done with the Feynman path integral method,\cite{FeynmanQMI}
which defines an amplitude called the kernel. The kernel
characterizes the trajectory of a particle starting from the point
$\alpha=(x_\alpha,y_\alpha,z_\alpha)$ at time $t_\alpha$ and
arriving a at the point $\beta=(x_\beta,y_\beta,z_\beta)$ at time
$t_\beta$. The kernel is a sum of all possible trajectories
between these two points and the times $t_\alpha$ and $t_\beta$.

Using the classical form of the Lagrangian
\begin{equation}
L(\dot{x},\dot{y},\dot{z},z,t)=m
\frac{{\dot{x}}^2}{2}+m\,\frac{{\dot{y}}^2}{2}+m
\frac{{\dot{z}}^2}{2}+mgz.
\end{equation}
Feynman\cite{FeynmanQMI} defined the kernel by
\begin{equation}
K(\beta,t_\beta;\alpha,t_\alpha)\sim\exp\big({\frac{i}{\hbar}S_{\rm
cl}(\beta,t_\beta;\alpha,t_\alpha)}\big) =
\exp\big(\frac{i}{\hbar}\!\int_{t_\alpha}^{t_\beta}L(\dot{x},\dot{y},
\dot{z},z,t)\,dt\big),
\end{equation}
with
$\int_{-\infty}^{+\infty}\int_{-\infty}^{+\infty}K(\beta,t_\beta;\alpha,t_\alpha)\,dx_\alpha\,dy_\alpha\,dz_\alpha=1$.
Hence,
\begin{equation}\label{eq:K}
K(\beta,t_\beta;\alpha,t_\alpha)=K_x(x_\beta,t_\beta;x_\alpha,t_\alpha)
K_y(y_\beta,t_\beta;y_\alpha,t_\alpha)
K_z(z_\beta,t_\beta;z_\alpha,t_\alpha),
\end{equation}
with
\begin{subequations}
\begin{eqnarray}
K_x(x_\beta,t_\beta;x_\alpha,t_\alpha)&=&\Big
(\frac{m}{2i\pi\hbar(t_\beta-t_\alpha)}\Big)^{1/2}
\exp{\frac{im}{\hbar}\Big(\frac{(x_\beta-x_\alpha)^2}{2(t_\beta-t_\alpha)}\Big)}
\label{eq:Kx}
\\
K_y(y_\beta,t_\beta;y_\alpha,t_\alpha)&=&\Big(\frac{m}{2i\pi\hbar(t_\beta-t_\alpha)}\Big)^{1/2}
\exp{\frac{im}{\hbar}\Big(\frac{(y_\beta-y_\alpha)^2}{2(t_\beta-t_\alpha)}\Big)}
\label{eq:Ky}
\\
K_z(z_\beta,t_\beta;z_\alpha,t_\alpha)&=&\Big(\frac{m}{2i\pi\hbar(t_\beta-t_\alpha)}\Big)^{1/2}
\exp{\frac{im}{\hbar}\Big(\frac{(z_\beta-z_\alpha)^2}{2(t_\beta-t_\alpha)}\Big)}.
\nonumber
\\
&&\times\exp{\frac{im}{\hbar}\Big(\frac{g}{2}(z_\beta+z_\alpha)(t_\beta-t_\alpha)-
\frac{g^2}{24}(t_\beta-t_\alpha)^3\Big)}. \label{eq:Kz}
\end{eqnarray}
\end{subequations}

For each atom with initial wave vector \textbf{k}, let us
designate by $\psi(\alpha,t_\alpha;\textbf{k} )$ the wave function
at time $t_\alpha$. We call S the set of points $\alpha$ where
this wave function does not vanish. It is then possible to
calculate the wave function at a later time $t_\alpha$ at points
$\beta$ such that there exits a straight line connecting $\alpha$
and $\beta$ for any point $\alpha \in S$. In this case,
Feynman\cite{FeynmanQMI} has shown that:
\begin{equation}\label{eq:psi}
 \psi(\beta,t;k_{0x},k_{0y},k_{0z}) =
\!\int_{(x_\alpha,y_\alpha,z_\alpha)\in
S}K(\beta,t;\alpha,t_\alpha)\,\psi(\alpha,t_\alpha;
k_{0x},k_{0y},k_{0z})\,dx_\alpha,dy_\alpha,dz_\alpha.
\end{equation}

For the double slit experiment, two steps are then necessary for
the calculation of the wave function: a first step before the
slits and a second step after the slits.

If we substitute Eqs.~(\ref{eq:psi0}) and (\ref{eq:K}) in
Eq.~(\ref{eq:psi}), we see that Feynman's path integral allows a
separation of variables, that is,
\begin{equation}
\psi(x,y,z,t;k_{0x},k_{0y},k_{0z}) =\psi_x(x,t;k_{0x})
\psi_y(y,t;k_{0y}) \psi_z(z,t;k_{0z}).
\end{equation}

References~\onlinecite{Shimizu} and \onlinecite{Tannoudji} treat
the vertical variable $z$ classically, which is shown in
Appendix~A to be a good approximation. Hence, we have
$z(t)=z_0+v_{0z}t+gt^2/2$. The arrival time of the wave packet at
the slits is $t_1(v_{0z},z_0)=\sqrt{\frac{2(l_1-z_0)}{g}+
\big(\frac{v_{0z}}{g}\big)^2}-\frac{v_{0z}}{g}$. For $v_{0z}=0$
and $z_0=0$, we have $t_1=\sqrt{\frac{2l_1}{g}}=124$\,ms and the
atoms have been accelerated to $v_{z1}=gt_{1}=1.22$\,m/s on
average at the slit. Thus the de Broglie wavelength
$\lambda=\hbar/mv_{z1} =1.8 \times 10^{-8}$\,m is two orders of
magnitude smaller than the slit width, $2\mu m$.

Because the two slits are very long compared with their other
dimensions, we will assume they are infinitely long, and there is
no spatial constraint on $y$. Hence, we have for an initial fixed
velocity $v_{0y}$:
\begin{equation}
\psi_y(y,t;k_{0y}) = \!\int_{y_\alpha}K_y(y,t;y_\alpha,t_\alpha=0)
\psi_0(y_\alpha;k_{0y})\,dy_\alpha.
\end{equation}
Thus
\begin{equation}
\label{eq:psiY} \psi_y(y,t;k_{0y})=(2\pi s_0^2(t))^{-1/4}
\exp[-\frac{(y- v_{0y}t)^2}{4\sigma_0 s_0(t)}+ik_{0y}(y-
v_{0y}t)],
\end{equation}
with $s_0(t)=\sigma_0(1+\frac{i\hbar t}{2m\sigma_0^2})$.

The wave packet is an infinite sum of wavepackets with fixed
initial velocity. The probability density as a function of $y$ is
\begin{equation}
\rho_y(y,t)
=\!\int_{-\infty}^{+\infty}(2\pi\tau^2)^{-1/2}e^{-k_y^2/2\tau^2}
|\psi_{y}(y,t;k_y)|^2\,dk_y =(2\pi\varepsilon_0^2(t))^{-1/2}\,
e^{-y^2/2\varepsilon_0^2(t)},
\end{equation}
with $\varepsilon_0^2(t)=\sigma_0^2(t)+(\frac{\hbar t\tau}{m})^2$
and $\sigma_0^2(t)=\sigma_0^2+ (\frac{\hbar t}{2m\sigma_0^2})^2$.
Because we know the dependence of the probability density on $y$,
in what follows we consider only the wave function
$\psi_x(x,t;k_{0x})$.

\subsection{The wave function before the slits}

\begin{figure}[h!]
\begin{center}
\includegraphics[width=0.9\linewidth]{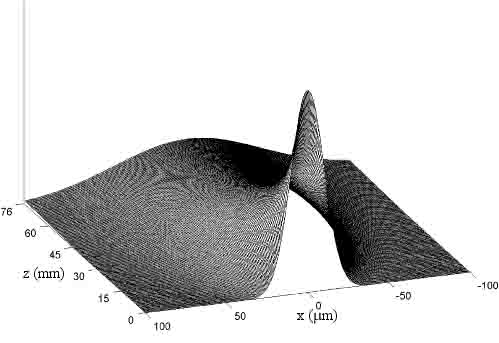}
\caption{\label{fig:ddpAvantFentes}Evolution of the density
$\rho_x(x,t)$ before the slits, The time of the scattering of the
initial wave packet ($t$ is determined from
$z=z_0+v_{0z}t+\frac{gt^2}{2}$).}
\end{center}
\end{figure}

Before the slits, we have
\begin{eqnarray}
\label{eq:psiXAvant} \psi_x(x,t;k_{0x}) &=& (2\pi
s_0^2(t))^{-1/4}\exp\big[-\frac{(x- v_{0x}t)^2}
{4\sigma_0 s_0(t)}+ik_{0x}(x-v_{0x}t)\big]\\
\rho_x(x,t)&=&(2\pi\varepsilon_0^2(t))^{-1/2}
\exp\big[-\frac{x^2}{2\varepsilon_0^2(t)}\big].
\end{eqnarray}
It is interesting that the scattering of the wave packet in $x$ is
caused by the dispersion of the initial position $\sigma_0$ and by
the dispersion $\tau$ of the initial velocity $v_{0x}$ (see
Fig.~\ref{fig:ddpAvantFentes}). Only 0.1\% of the atoms will cross
through one of the slits; the others will be stopped by the plate.

\subsection{The wave function after the slits}

\begin{figure}[h!]
\begin{center}
\includegraphics[width=0.7\linewidth]{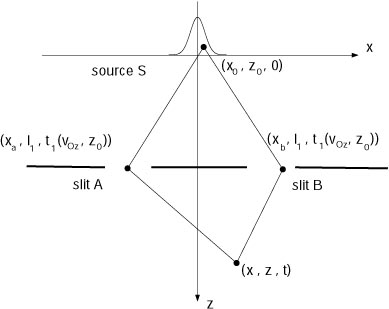}
\caption{\label{fig:explicationCalculs}Schematic representation of
the experiment: calculation method of the wave function after the
slits.}
\end{center}
\end{figure}

The wave function after the slits with fixed $z_0$ and
$k_{0z}=mv_{0z}/\hbar$ for $t\geq t_1(v_{0z},z_0)$ is deduced from
the values of the wave function at slits A and B
(cf.Fig.~\ref{fig:explicationCalculs}) by using
Eq.~(\ref{eq:psi}). We obtain:
\begin{equation}
\label{eq:psiXApres} \psi_x(x,t;k_{0x},k_{0z},z_{0}) = \psi_A +
\psi_B,
\end{equation}
with
\begin{subequations}
\label{14}
\begin{eqnarray}
\psi_A=\!\int_{A}K_x(x,t;x_a,t_1(v_{0z},z_{0}))\,
\psi_x(x_a,t_1(v_{0z},z_{0});k_{0x})\, dx_a \label{eq:psiA}
\\
\psi_B= \!\int_{B}K_x(x,t;x_b,t_1(v_{0z},z_{0}))\,\psi_x(x_b,t_1
(v_{0z},z_{0});k_{0x}) \,dx_b, \label{eq:psiB}
\end{eqnarray}
\end{subequations}
where $\psi_x(x_a,t_1(v_{0z},z_{0});k_{0x})$ and $\psi_x(x_b,t_1
(v_{0z},z_{0});k_{0x})$ are given by Eq.~(\ref{eq:psiXAvant})
whereas $K_x(x,t;x_a,t_1(v_{0z},z_{0}))$ and
$K_x(x,t;x_b,t_1(v_{0z},z_{0}))$ are given by Eq. (\ref{eq:Kx}).

The probability density is
\begin{equation}
\label{eq:roXApres}
\rho_{x}(x,t;k_{0z},z_{0})=\!\int_{-\infty}^{+\infty}(2\pi\tau^2)^{-1/2}
\exp\Big(-\frac{k_{0x}^2}{2\tau^2}\Big)|
\psi_x(x,t;k_{0x},k_{0z},z_{0})|^2\,dk_{0x}.
\end{equation}
The arrival time $t_2$ of the center of the wave packet on the
detecting plate depends on $z_{0}$ and $v_{0z}$. We have
$t_2=\sqrt{\frac{2(l_1+l_2-z_{0})}{g}+
\big(\frac{v_{0z}}{g}\big)^2} -\frac{v_{0z}}{g}$. For $z_{0}=0$
and $v_{0z}=0$, $t_2=\sqrt{\frac{2(l_1+l_2)}{g}}=196$\,ms and the
atoms are accelerated to $v_{z2}=g t_{2}=1.93$\,m/s.

\begin{figure}[h!]
\begin{center}
\includegraphics[width=0.8\linewidth]{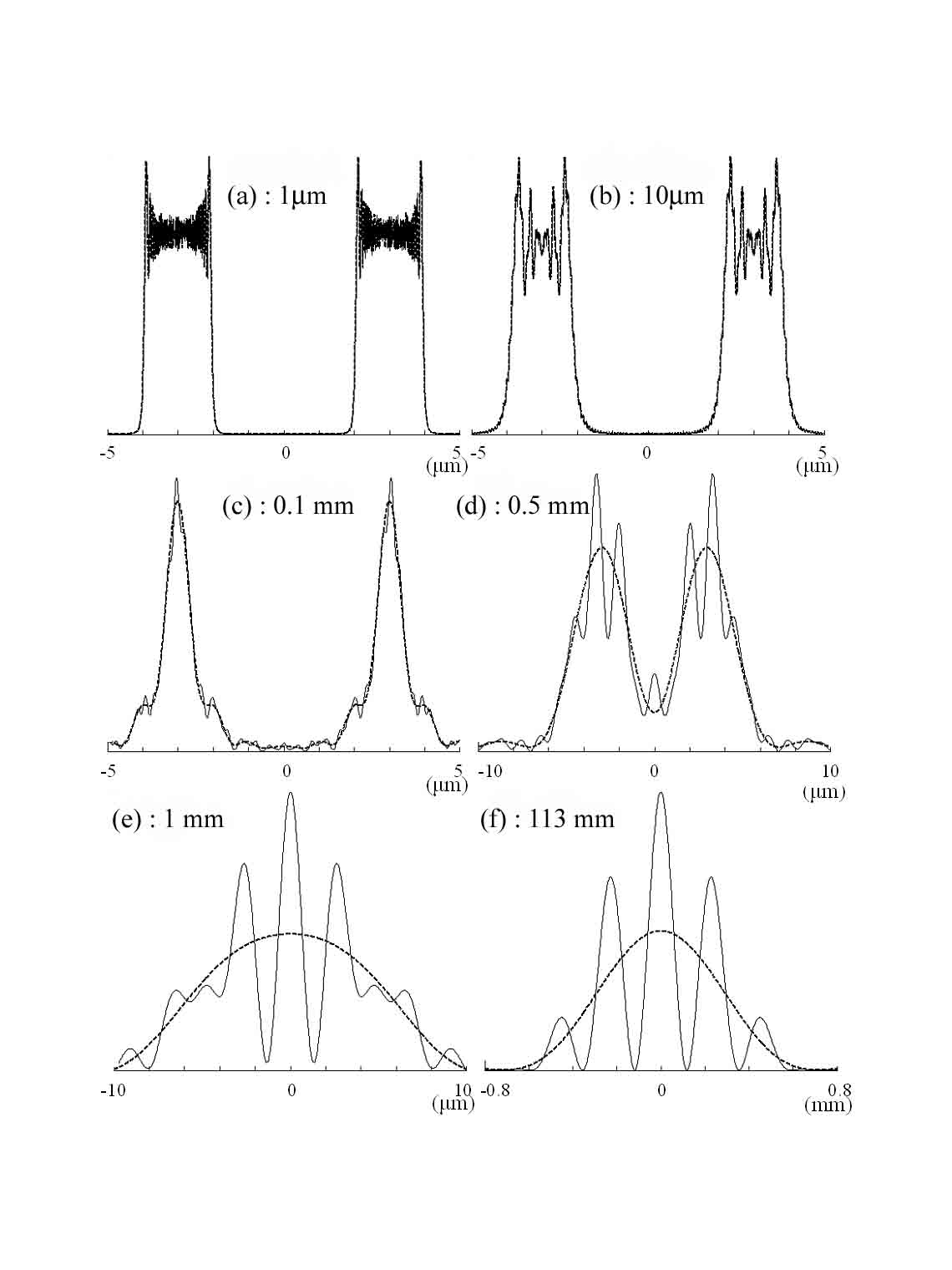}
\caption{\label{fig:comparaisonInterferenceDiffraction}Comparison
between $|\psi_A+\psi_B|^2$ (plain line) and
$|\psi_A|^2+|\psi_B|^2$ (dotted line) with $z_{0}=0$ and
$k_{0z}=0$ at (a) $1\,\mu$m, (b) $10\,\mu$ m, (c) 0.1\,mm, (d)
0.5\,mm, (e) 1\,mm, and (f) 113\,mm after the slits.}
\end{center}
\end{figure}

The calculation of $\rho_{x}(x,t;k_{0z},z_{0})$ at any $(x,t)$
with $k_{0z}$ and $z_{0}$ given and $t\geq t_1$ is done by a
double numerical integration: (a) Eq.~(\ref{eq:roXApres}) is
integrated numerically using a discretization of $k_{0x}$ into 20
values; (b) the integration of Eq.~(\ref{eq:psiXApres}) using
Eqs.~(\ref{14}) is done by a discretization of the slits $A$ and
$B$ into 200 values each.
Figure~\ref{fig:comparaisonInterferenceDiffraction} shows the
cross sections of the probability density ($|\psi_A + \psi_B|^2$)
for $z_{0}=0$, $v_{0z}=0$ $(k_{0z}=0)$ and for several distances
($\Delta z=\frac{1}{2}gt^2 -\frac{1}{2}gt_1^2$) after the double
slit: $1\,\mu$m, $10\,\mu$ m, 0.1\,mm, 0.5\,mm, 1\,mm and 113\,mm.

\begin{figure}[h!]
\begin{center}
\includegraphics[width=0.9\linewidth]{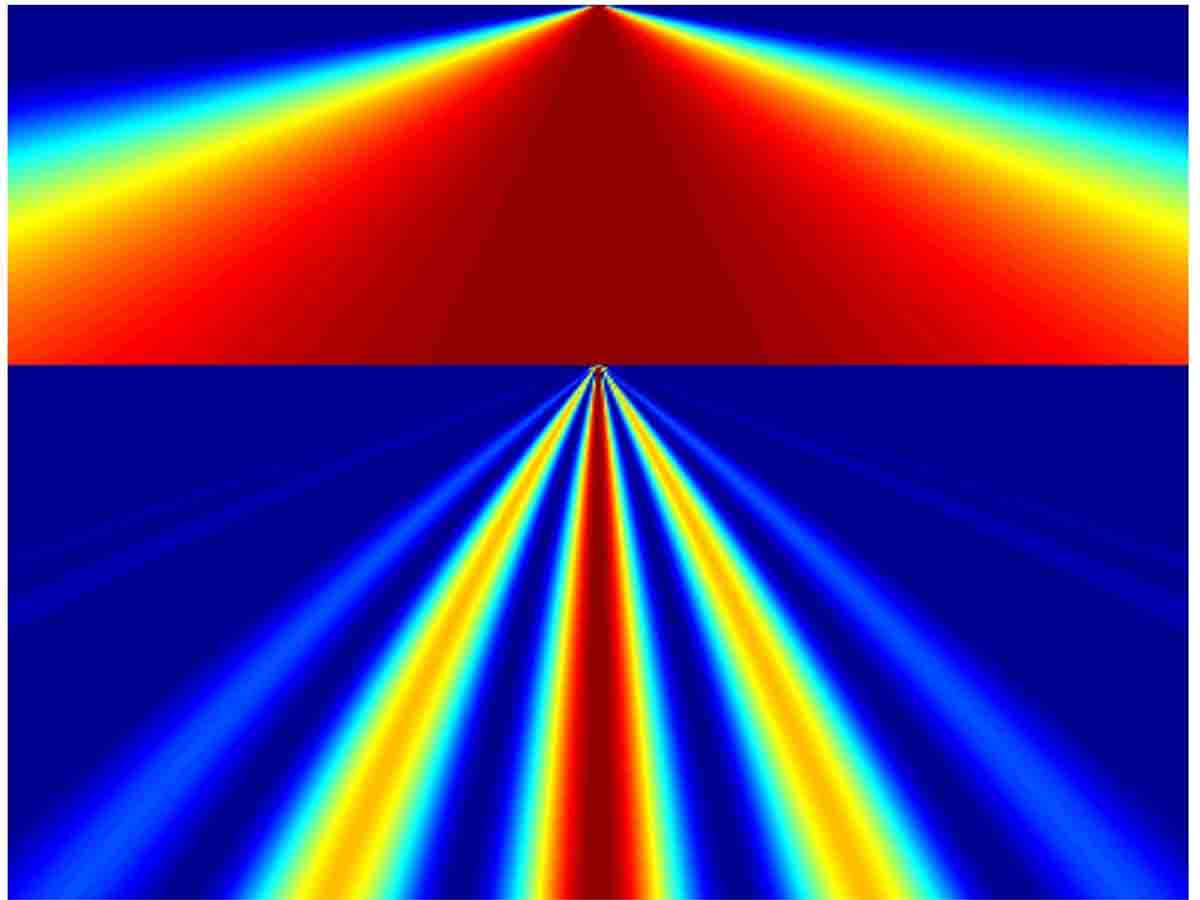}
\caption{\label{fig:ddp3Dtotal}Evolution of the probability
density $\rho_x(x,t;k_{0z}=0,z_{0}=0)$ from the source to the
detector screen.}
\end{center}
\end{figure}

\begin{figure}[h!]
\begin{center}
\includegraphics[width=0.8\linewidth]{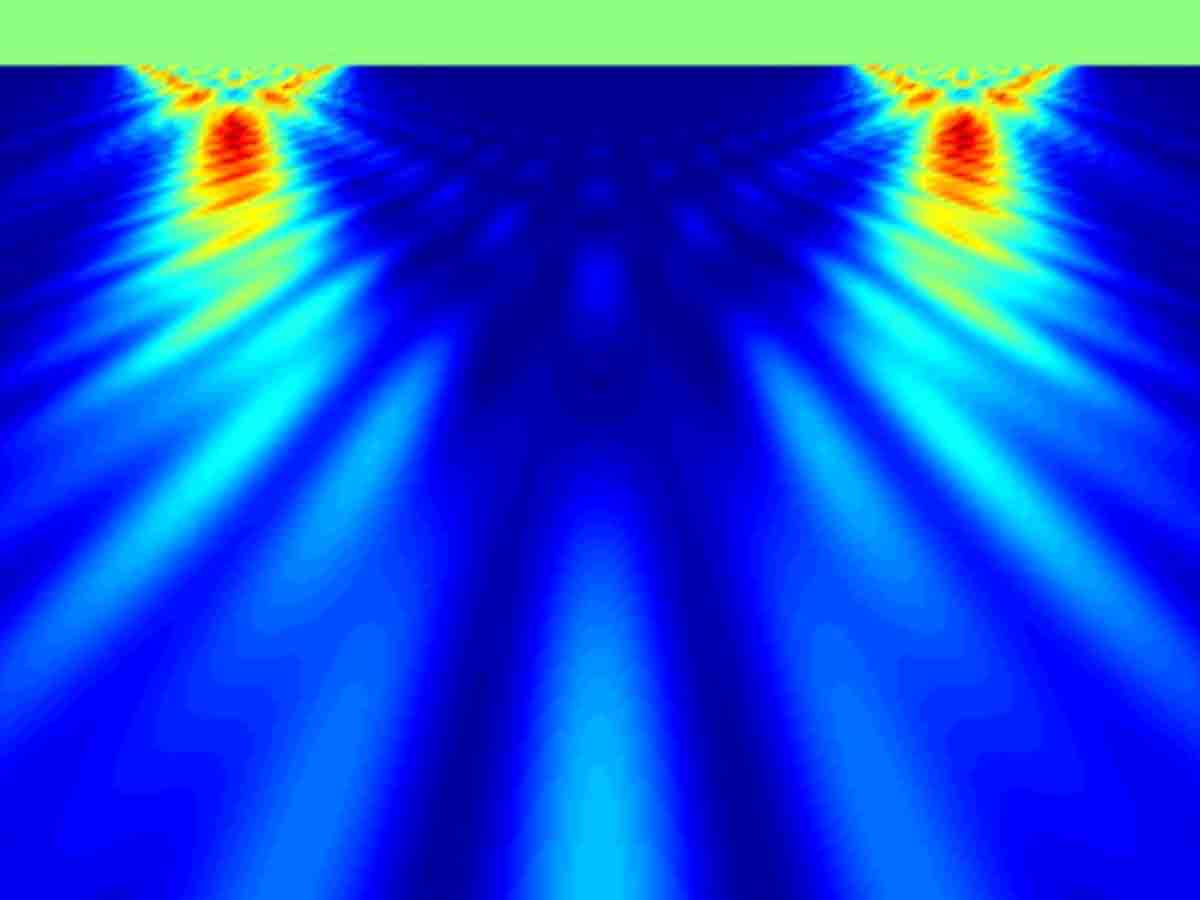}
\caption{\label{fig:ddp3D_1e-3}Evolution of the probability
density $\rho_x(x,t;k_{0z}=0,z_{0}=0)$ for the first millimeter
after the slits.}
\end{center}
\end{figure}

\begin{figure}[h!]
\begin{center}
\includegraphics[width=0.8\linewidth]{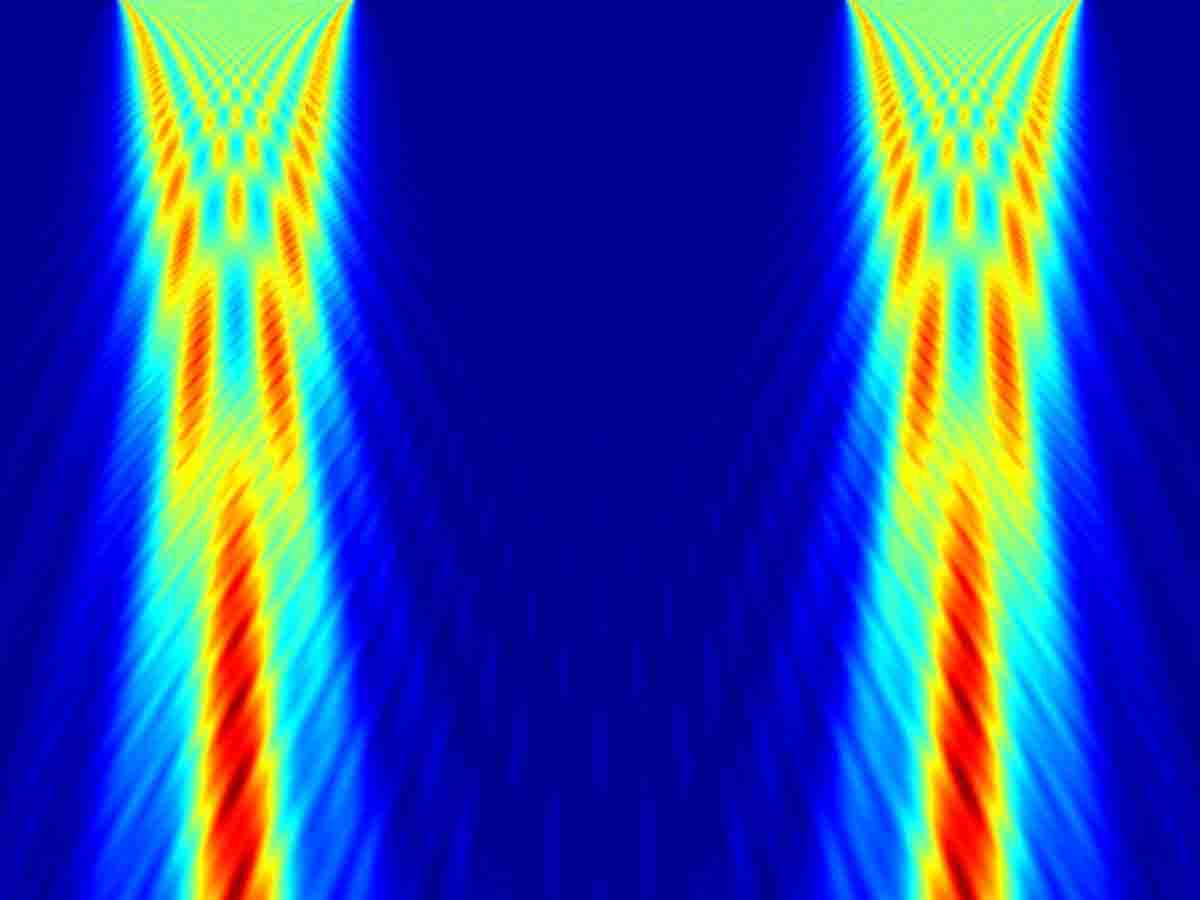}
\caption{\label{fig:ddp3D_1e-4}Evolution of the probability
density $\rho_x(x,t;k_{0z}=0,z_{0}=0)$ for the first 100\,$\mu$m
after the slits.}
\end{center}
\end{figure}

The calculation method enables us to compare the evolution of the
probability density when both slits are simultaneously open
(interference: $|\psi_A+\psi_B|^2$) with the sum of the evolutions
of the probability density when the two slits are successively
opened (sum of two diffraction phenomena:
($|\psi_A|^2+|\psi_B|^2$).
Figure~\ref{fig:comparaisonInterferenceDiffraction} shows the
probability density ($|\psi_A|^2+|\psi_B|^2$) for the same cases.
Note that the difference between the two phenomena does not exist
immediately at the exit of the two slits; differences appear only
after some millimeters after the slits.

Figures~5--7 show the evolution of the probability density. At
0.1\,mm after the slits, we know through which slit each atom has
passed, and thus the interference phenomenon does not yet exist
(see Figs.~\ref{fig:comparaisonInterferenceDiffraction} and
\ref{fig:ddp3D_1e-4}). Only at 1\,mm after the slits do the
interference fringes become visible, just as we would expect by
the Fraunhoffer approximation (see
Figs.~\ref{fig:comparaisonInterferenceDiffraction} and
\ref{fig:ddp3D_1e-3}).

\subsection{Comparison with the Shimizu experiment}

In the Shimizu experiment, atoms arrive at the detection screen
between $t=t_{\min}$ and $t_{\max}$. To obtain the measured
probability density in this time interval, we have to sum the
probability density above the initial position $z_{0}$ and their
initial velocity $v_{0z}$ compatible with $t_{\min}\leq t_{2}\leq
t_{\max}$, that is,
\begin{equation}
\rho_{x}(x,t_{\min}\leq t\leq t_{\max})=\!\int_{t_{\min}\leq t_{2}
\leq t_{\max}} \hspace{-1cm}
\rho_{x}(x,t_{2};k_{0z},z_{0})\,e^{-k_{0z}^{2}/2 \tau^{2}}
e^{-z_{0}^2/2 \sigma_{z}^2}\, dk_{0z} dz_{0}.
\end{equation}
The positions at the detection screen can only be measured to
about 80\,$\mu$m, and thus to compare our results with the
measured results. we perform the average
\begin{equation}
\rho_{\rm measured}(x,t_{\min}\leq t\leq
t_{\max})=\frac{1}{80\,\mu \rm m}\!\int_{x-40 \mu m}^{x+40 \mu m}
\rho(u,t_{\min}\leq t\leq t_{\max})\,du.
\end{equation}

\begin{figure}[h!]
\begin{center}
\includegraphics[width=0.9\linewidth]{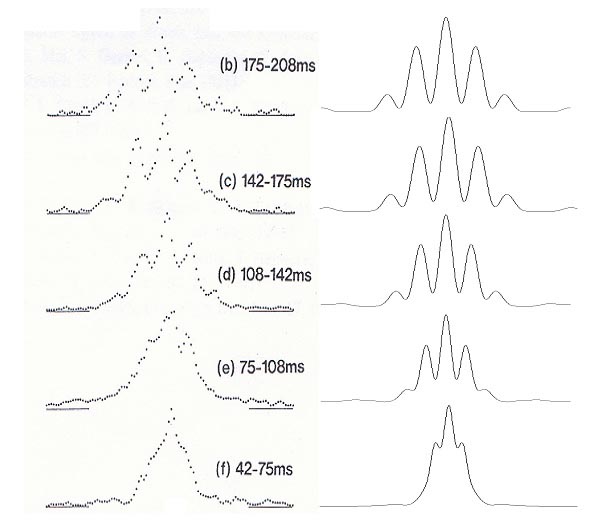}
\caption{\label{fig:comparaisonImpactsMesure}Comparison of the
probability density measured experimentally by
Shimizu\cite{Shimizu} (left) and the probability density
calculated numerically with our model (right).}
\end{center}
\end{figure}

Figure~\ref{fig:comparaisonImpactsMesure} compares those
calculations to the results found in Ref.~\onlinecite{Shimizu}.
The experimental fringe separation is narrower than in our
calculation, see Figs.~\ref{fig:comparaisonImpactsMesure} and
\ref{fig:impacts}. This difference is explained by a technical
problem in the Shimizu experiment.\cite{Shimizu}

\section{Impacts on screen and trajectories}

In the Shimizu experiment the interference fringes are observed
through the impacts of the neon atoms on a detection screen. It is
interesting to simulate the neon atoms trajectories in the de
Broglie-Bohm interpretation,\cite{deBroglie,Bohm} which accounts
for atom impacts. In this formulation of quantum mechanics, the
particle is represented not only by its wave function, but also by
the position of its center of mass. The atoms have trajectories
which are defined by the speed $\textbf{v}(x,y,z,t)$ of the center
of mass, which at position $(x,y,z)$ at time $t$ is given
by\cite{Holland,GondranMA}
\begin{equation}
\label{eq:vitesse} \textbf{v}(x,y,z,t) = \frac{\nabla
S(x,y,z,t)}{m} + \frac{\nabla\log\rho(x,y,z,t)
\times\textbf{s}}{m}
 = \frac{\hbar}{m\rho} [{\rm Im}(\psi^*\nabla\psi)+{\rm
Re}(\psi^*\nabla\psi) \times \frac{\textbf{s}}{|\textbf{s}|}],
\end{equation}
where
$\psi(x,y,z,t)=\sqrt{\rho(x,y,z,t)}\exp{\left(\frac{i}{\hbar}
S(x,y,z,t)\right)}$ and $\textbf{s}$ is the spin of the particle.
Let us see how this interpretation gives the same experimental
results as the Copenhagen interpretation.

If $\psi$ satisfies the Schr\"odinger equation,
\begin{equation}
\label{eq:schrodinger} i\hbar \frac{\partial \psi}{\partial t}=-
\frac{\hbar^2}{2m} \nabla^2\psi +V\psi,
\end{equation}
with the initial condition
$\psi(x,y,z,0)=\psi_{0}(x,y,z)=\sqrt{\rho_{0}(x,y,z)}
\exp(\frac{i}{\hbar} S_{0}(x,y,z))$, then $\rho$ and $S$ satisfy:
\begin{eqnarray}
\label{eq:HJ} \frac{\partial S}{\partial t}+\frac{1}{2m}(\nabla
S)^{2}+V -\frac{\hbar^2}{2m}\frac{\triangle \sqrt{\rho}}{
\sqrt{\rho}}&=&0\\
\label{eq:continuite} \frac{\partial \rho}{\partial t} + \nabla
\cdot \Big(\rho \frac{\nabla S }{m}\Big) &=& 0,
\end{eqnarray}
with initial conditions $S(x,y,z,0)=S_{0}(x,y,z)$ and
$\rho(x,y,z,0)=\rho_{0}(x,y,z)$.

In both interpretations, $\rho(x,y,z,t)=|\psi(x,y,z,t)|^{2}$ is
the probability density of the particles. But, in the Copenhagen
interpretation, it is a postulate for each $t$ (confirmed by
experience). In the de Broglie-Bohm interpretation, if
$\rho_0(x,y,z)$ is the probability density of presence of
particles for $t=0$ only, then $\rho(x,y,z,t)$ must be the
probability density of the presence of particles without any
postulate because Eq.~(\ref{eq:continuite}) becomes the Madelung
equation:
\begin{equation}
\frac{\partial \rho}{\partial t} + \nabla \cdot (\rho\textbf{v}) =
0,
\end{equation}
(thanks to $\textbf{v}= \frac{\nabla S}{m}+\nabla \times
(\frac{\ln \rho}{m}) \textbf{s}$), which is obviously the fluid
mechanics equation of conservation of the density. The two
interpretations therefore yield statistically identical results.
Moreover, the de Broglie-Bohm theory naturally explains the
individual impacts.

In the initial de Broglie-Bohm
interpretation,\cite{deBroglie,Bohm} which was not relativistic
invariant, the velocity was not given by Eq.~(\ref{eq:vitesse}),
but by $\textbf{v}=\frac{\nabla S}{m}$ which does not involve the
spin. In the Shimizu experiment, the spin of each neon atom in the
magnetic trap was constant and vertical:
$\textbf{s}=(0,0,\frac{\hbar}{2})$. In our case the spin-dependent
term
$\frac{\nabla\log\rho}{m}\times\textbf{s}=\frac{\hbar}{2m\rho}
\big(\frac{\partial\rho}{\partial y},-\frac{\partial\rho}{\partial
x},0\big)$ is negligible after the slit, but not before.

For the simulation, we choose at random (from a normal
distribution $f(0,0,0;\sigma_k,\sigma_k,\sigma_k)$ the wave vector
$\textbf{k}=(k_{0x},k_{0y},k_{0z})$ to define the initial wave
function (\ref{eq:psi0}) of the atom prepared inside the
magneto-optic trap. For the de Broglie-Bohm interpretation, we
also choose at random the initial position $(x_0,y_0,z_0)$ of the
particle inside its wave packet (normal distribution
$f((0,0,0);(\sigma_0,\sigma_0,\sigma_z))$). The trajectories are
given by
\begin{subequations}
\begin{eqnarray}
\frac{dx}{dt}&=&v_x(x,t) =\frac{1}{m}\frac{\partial S}{\partial x}
+\frac{\hbar}{2m\rho}\frac{\partial\rho}{\partial y}
\\
\frac{dy}{dt}&=&v_y(x,t) =\frac{1}{m}\frac{\partial S}{\partial y}
-\frac{\hbar}{2m\rho}\frac{\partial\rho}{\partial x}
\\
\frac{dz}{dt}&=&v_z(x,t) =\frac{1}{m}\frac{\partial S}{\partial
z},
\end{eqnarray}
\end{subequations}
where
$\rho(x,y,t;k_{0x},k_{0y})=|\psi_{x}(x,t;k_{0x})\psi_{y}(y,t;k_{0y})|^{2}$
and $\psi_{x}$ and $\psi_{y}$ are given by
Eqs.~(\ref{eq:psiY})--(\ref{eq:psiXApres}).

\subsection{Trajectories before the slits}

Before the slits, Appendix~B gives $z(t)
=z_0\frac{\sigma_z(t)}{\sigma_z}+ v_{0z}t + \frac{1}{2}gt^2$,
$x(t) = v_{0x}t + \sqrt{x_0^2+y_0^2} \frac{\sigma_0(t)}{\sigma_0}
\cos{\varphi(t)}$, and $y(t) = v_{0y}t +
\sqrt{x_0^2+y_0^2}\frac{\sigma_0(t)}{\sigma_0} \sin{\varphi(t)}$,
with $\varphi(t)=\varphi_0+{\rm arctan}\big(-\frac{\hbar
t}{2m\sigma_0^2}\big)$, $\cos
\varphi_0=\frac{x_0}{\sqrt{x_0^2+y_0^2}}$ and $\sin
\varphi_0=\frac{y_0}{\sqrt{x_0^2+y_0^2}}$. For a given wave vector
$\textbf{k}$ and an initial position $(x_0,y_0,z_0)$ inside the
wave packet, an atom of neon will arrive at a given position on
the plate containing the slits. Notice that the term
$\nabla\log\rho\times\textbf{s}/m$ adds to the trajectory defined
by $\nabla S/m$ a rotation of $-\pi/2$ around the spin axis (the
$z$ axis).

\begin{figure}[h!]
\begin{center}
\includegraphics[width=0.9\linewidth]{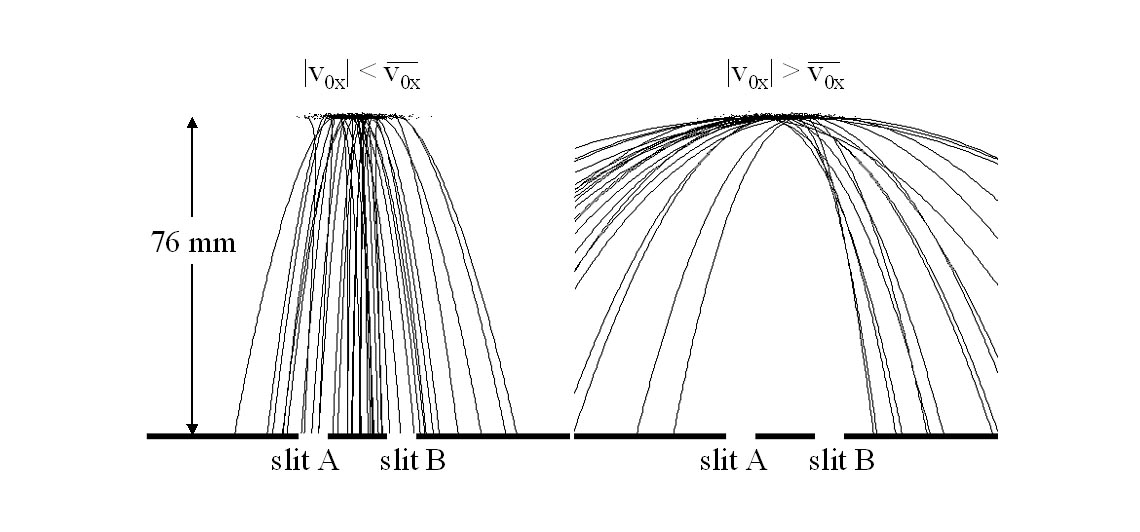}
\caption{\label{fig:trajectoiresAvantFentes}Trajectories of atoms
before the slit. Note that if the initial velocities $|v_{0x}|\geq
3.9\times 10^{-4}$\,m/s, then no atoms can cross the slit.}
\end{center}
\end{figure}

The source atoms do not all pass through the slits; most of them
are stopped by the plate. Only atoms having a small horizontal
velocity $v_{0x}$ can go through the slits. Indeed an atom with an
initial velocity $v_{0x}$ and an initial position $x_0,y_0$
arrives at the slits at $t=t_1$ at the horizontal position
$x(t_1)=v_{0x}t_1+y_0 (\sigma_0(t_1)/\sigma_0)$. For this atom to
go through one of the slits, it is necessary that $|x(t_1)|\leq
\overline{x}$, with $\overline{x}=(d+b)/2=4\times 10^{-6}$\,m and
$-2\sigma_0\leq y_0\leq 2\sigma_0$ and $t_1\simeq 0.124$\,s.
Consequently, it is necessary that the initial velocities of the
atoms satisfy $|v_{0x}|\leq (\overline{x}+2\sigma_0(t_1))/t_1
=\overline{v}_{0x}\simeq 3.9\times 10^{-4}$\,m/s. The double slit
filters the initial horizontal velocities and transforms the
source atoms after the slits into a quasi-monochromatic source.
The horizontal velocity of an atom leads to a horizontal shift of
the atom's impacts on the detection screen. The maximum shift is
$\Delta x=\overline{v}_{0x}\Delta t$, where $\Delta t$ is the time
for the atom to go from the slits to the screen ($\Delta
t=t_2-t_1\simeq 0.072$\,s); hence $\Delta x\simeq 2.8\times
10^{-5}$\,m. This shift does not produce a blurring of the
interference fringes because the interference fringes are
separated from one another by $25\times 10^{-5}\,{\rm m}\gg \Delta
x$. Note that if the source was nearer to the double slit (for
example if $l_1=5$\,mm, then $\Delta x\simeq 10\times
10^{-5}$\,m), the slit would not filter enough horizontal
velocities and consequently the interference fringes would not be
visible.

The system appears fully deterministic. If we know the position
and the velocity of an atom inside the source, then we know if it
can go through the slit or not.
Figure~\ref{fig:trajectoiresAvantFentes} shows some trajectories
of the source atoms as a function of their initial velocities.
Only atoms with a velocity $|v_{0x}|\leq \overline{v}_{0x}$ can go
through the slits.

\subsection{Velocities and trajectories after the slits}

\begin{figure}[h!]
\begin{center}
\includegraphics[width=0.9\linewidth]{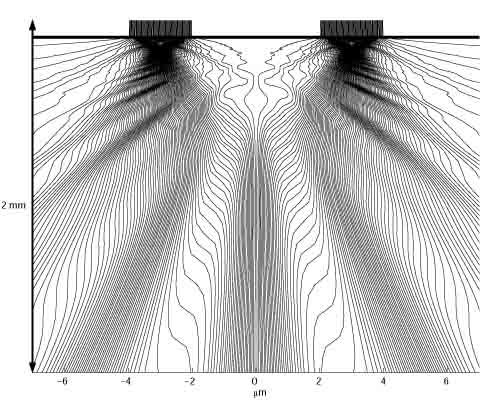}
\caption{\label{fig:zoomTrajectoires}Zoom of trajectories of atoms
for the first millimeters after the slit ($k_{0x}=k_{0z}=0$).}
\end{center}
\end{figure}

In what follows, we consider only atoms that have gone through one
of the slits. After the slits, we still have $z(t)=v_{0z}t
+\frac{1}{2}gt^2+z_0(\sigma_z(t)/\sigma_z)$, but now $v_x(t)$ and
$v_y(t)$ and $x(t)$ and $y(t)$ have to be calculated numerically.
The calculation of $v_x(x,t)$ is done by a numerical computation
of an integral in $x$ above the slits A and B (see Appendix B);
$x(t)$ is calculated with a Runge-Kutta method.\cite{Lambert} We
use a time step $\Delta t$ which is inversely proportional to the
acceleration. At the exit of the slit, $\Delta t$ is very small:
$\Delta t \simeq 10^{-8}$\,s; it increases to $\Delta t \simeq
10^{-4}$\,s at the detection screen.
Figure~\ref{fig:zoomTrajectoires} shows the trajectories of the
atoms just after the slits; $x_0$ and $y_0$ are drawn at random,
$z_0=0$, with $v_{0x}=v_{0z}=0$.

\subsection{Impacts on the screen}

\begin{figure}[h!]
\begin{center}
\includegraphics[width=0.9\linewidth]{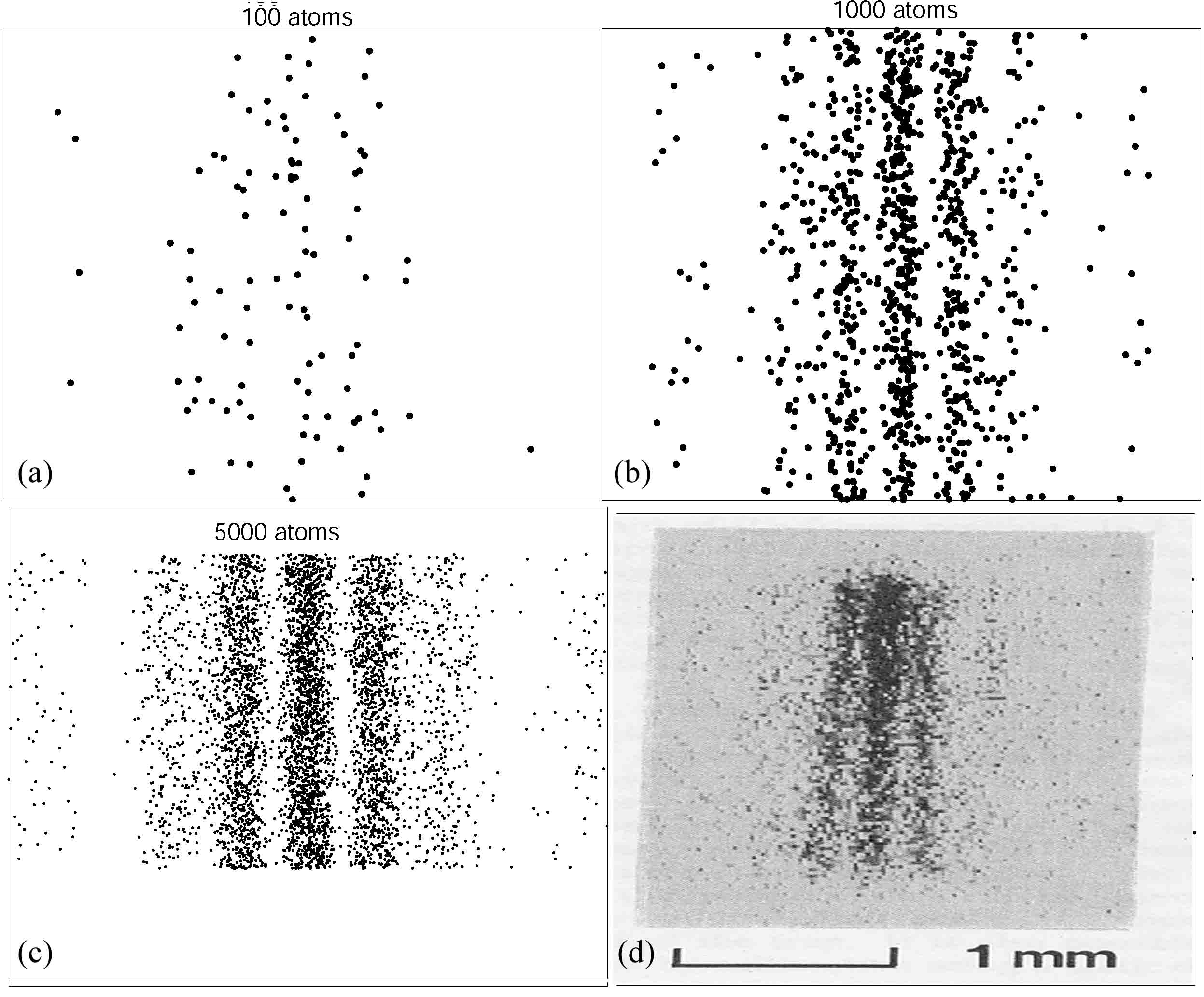}
\caption{\label{fig:impacts}Atomic impacts on the screen of
detection.}
\end{center}
\end{figure}

We observe the impact of each particle on the detection screen as
shown by the last image in Fig.~\ref{fig:impacts}. The classic
explanation of these individual impacts on the screen is the
reduction of the wave packet. An alternative interpretation is
that the impacts are due to the decoherence caused by the
interaction with the measurement apparatus.

In the de Broglie-Bohm formulation of quantum mechanics, the
impact on the screen is the position of the center of mass of the
particle, just as in classical mechanics. Figure~\ref{fig:impacts}
shows our results for 100, 1000, and 5000 atoms whose initial
position $(x_0,y_0,z_0)$ are drawn at random. The last image
corresponds to 6000 impacts of the Shimizu
experiment.\cite{Shimizu} The simulations show that it is possible
to interpret the phenomena of interference fringes as a
statistical consequence of a particle trajectories.

\section{Summary}

We have discussed a simulation of the double slit experiment from
the source of emission, passing through a realistic double slit,
and its arrival at the detector. This simulation is based on the
solution of Schr\"odinger's equation using the Feynman path
integral method. A simulation with the parameters of the 1992
Shimizu experiment produces results consistent with their
observations. Moreover the simulation provides a detailed
description of the phenomenon in the space just after the slits,
and shows that interference begins only after 0.5\,mm. We also
show that it is possible to simulate the trajectories of particles
by using the de Broglie-Bohm interpretation of quantum mechanics.

\appendix

\section{Calculation of $\psi_z(z,t;k_{0z})$}

Because there is no constraints on the vertical variable $z$, we
find using Eqs.~(\ref{eq:Kz}) and (\ref{eq:psi}) for all $t>0$
(before and after the double slit) that
\begin{equation}
\psi_z(z,t;k_{0z}) = \!\int_S\!
K_z(z,t;z_\alpha,t=0)\times\psi_{0_z}(z_\alpha;k_{0z})\,dz_\alpha,
\end{equation}
where the integration is done over the set S of the points
$z_\alpha$  where the initial wave packet
$\psi_{0_z}(z_\alpha;k_{0z})$ does not vanish. We obtain:
\begin{eqnarray}
\psi_z(z,t;k_{0z}) &= &(2\pi s_z^2(t))^{-1/4}
\exp\Big(-\frac{(z-v_{0z}t-gt^2/2)^2}{4\sigma_z s_z(t)}\Big)
\nonumber
\\
{}&&\times \exp\Big[\frac{i m}{\hbar}\big((v_{0z}+g
t)(z-v_{0z}t/2)-\frac{mg^2t^3}{6}\big)\Big], \label{eq:psiZ}
\end{eqnarray}
where $s_z(t)=\sigma_z\big(1+\frac{i\hbar t}{2m\sigma_z^2}\big)$.
Consequently we have:
\begin{equation}
|\psi_z(z,t;v_{0z})|^2=(2\pi\sigma_z^2(t))^{-1/2}\exp\Big[-
\frac{(z-v_{0z}t-gt^2/2)^2}{2\sigma_z^2(t)}\Big],
\end{equation}
with $\sigma_z(t)=|s_z(t)|=\sigma_z\big(1+\big(\frac{\hbar
t}{2m\sigma_z^2}\big)^2\big)^{1/2}$.

Note that $\sigma_z(t)$ is negligible compared to $l_1$
($\sigma_z=0.3\,$mm and $\frac{\hbar t}{2m\sigma_z}=10^{-3}$\,mm
are negligible compared to $l_1=76$\,mm for an average crossing
time inside the interferometer of $t\sim 200$\,ms). Therefore
$(2\pi\sigma_z^2(t))^{-1/2}\exp\big[-\frac{(z-v_{0z}t-gt^2/2)^2}
{2\sigma_z^2(t)}\big]\simeq\delta_0(z-v_{0z}t-\frac{gt^2}{2})$,
and if $z_{0}$ is the initial position of the particle, we have $z
\simeq z_{0}+v_{0z}t+\frac{gt^2}{2}$ at time $t$.

\section{Calculation of the atom's trajectories}

The velocity (\ref{eq:vitesse}) applied to Eq.~(\ref{eq:psiZ})
gives the differential equation for the vertical variable $z$,
\begin{equation}
\frac{dz}{dt}=v_z(z,t) =\frac{1}{m}\frac{\partial S}{\partial z}
=v_{0z} +gt
+\frac{(z-v_{0z}t-gt^2/2)\hbar^2t}{4m^2\sigma_z^2\sigma_z^2(t)},
\end{equation}
from which we find
\begin{equation}
\label{exactly} z(t)=v_{0z}t
+\frac{1}{2}gt^2+z_0\frac{\sigma_z(t)}{\sigma_z}.
\end{equation}
Equation~(\ref{exactly}) gives the classical trajectory if $z_0=0$
(the center of the wave packet).

The velocity (\ref{eq:vitesse}) applied before the slit to
Eqs.~(\ref{eq:psiY}) and (\ref{eq:psiXAvant}) gives the
differential equations in the $x$ and $y$ directions:
\begin{subequations}
\begin{eqnarray}
 \frac{dx}{dt}&=&v_x(x,t) =\frac{1}{m}\frac{\partial S}{\partial
x} +\frac{\hbar}{2m\rho}\frac{\partial\rho}{\partial y} =v_{0x}
+\frac{(x-v_{0x}t)\hbar^2t}{4m^2\sigma_0^2\sigma_0^2(t)}
-\frac{\hbar(y-v_{0y}t)}{2m\sigma_0^2(t)}
\\
\frac{dy}{dt}&=&v_y(x,t) =\frac{1}{m}\frac{\partial S}{\partial y}
-\frac{\hbar}{2m\rho}\frac{\partial\rho}{\partial x} =v_{0y}
-\frac{(y-v_{0y}t)\hbar^2t}{4m^2\sigma_0^2\sigma_0^2(t)}
+\frac{\hbar(x-v_{0x}t)}{2m\sigma_0^2(t)}.
\end{eqnarray}
\end{subequations}
It then follows that
\begin{subequations}
\begin{eqnarray}
x(t)&=&v_{0x}t+\sqrt{x_0^2+y_0^2} \frac{\sigma_0(t)}{\sigma_0}
\cos{\varphi(t)}
\label{eq:xAvanta} \\
 y(t)&=&v_{0y}t+\sqrt{x_0^2+y_0^2}
\frac{\sigma_0(t)}{\sigma_0} \sin{\varphi(t)} \label{eq:xAvantb},
\end{eqnarray}
\end{subequations}
with $\varphi(t)=\varphi_0+{\rm arctan}\big(-\frac{\hbar
t}{2m\sigma_0^2}\big)$,
$\cos(\varphi_0)=\frac{x_0}{\sqrt{x_0^2+y_0^2}}$, and
$\sin(\varphi_0)=\frac{y_0}{\sqrt{x_0^2+y_0^2}}$.
Equations~(\ref{eq:xAvanta}) and (\ref{eq:xAvantb}) give the
classical trajectory if $x_0=y_0=0$ (the center of the wave
packet).

After the slits, the velocity $v_x(x,t) = \frac{\hbar}{m}
\frac{Im( \frac{\partial \psi}{\partial x}\psi^*)}{\psi\psi^*}$
given by Eq. (\ref{eq:vitesse}) can be calculated using Eqs.
 (\ref{eq:psi}), (\ref{eq:psiA}) and (\ref{eq:psiB}). We obtain:

 \begin{equation*}
 v_x(x,t)= \frac{1}{t-t_1}\left[x
+\frac{-1}{2(\alpha^2+\beta_t^2)}\left(\beta_t
Im\left(\frac{C(x,t)}{H(x,t )}\right) + \alpha
Re\left(\frac{C(x,t)}{H(x,t)}\right) - \beta_t \gamma_{x,t}
\right) \right]
\end{equation*}
with
\begin{eqnarray*}
H(x,t) = \int_{X_A - b}^{X_A + b}f(x,u,t)\,du + \int_{X_B -
b}^{X_B + b}f(x,u,t)\,du
\\
C(x,t) = \left[f(x,u,t)\right]_{u= X_A - b}^{u= X_A + b} +
\left[f(x,u,t)\right]_{u= X_B - b}^{u= X_B + b}
\end{eqnarray*}
where $X_A$ and $X_B$ are the centers of the two slits, and where
\begin{eqnarray*}
                                    &&f(x,u,t) = \exp{[(\alpha+i\beta _t)u^2+i\gamma _{x,t}u]}
                                    \\
                                    &&\alpha = - \frac{1}{4\sigma_0^2\left(1+\left(\frac{\hbar t_1}{2m\sigma_0^2}\right)^2\right)}
                                    \\
                                    &&\beta _t = \frac{m}{2\hbar}\left(\frac{1}{t-t_1}+\frac{1}{t_1\left(1+\left(\frac{2m\sigma_0^2}{\hbar t_1}\right)^2\right)}\right)
                                    \\
                                    &&\gamma _{x,t} = -\frac{mx}{\hbar (t-t_1)}.
\end{eqnarray*}

\section{Convergence}

In simulation, it is possible to make the Planck constant $h$ tend
towards zero. So, it is possible to show, about the example of the
Young slits, the convergence of the quantum mechanics to the
classical mechanics. In the following simulations, one take
$k_{0x}=k_{0z}=0$.

\begin{figure}[h!]
\begin{center}
\includegraphics[width=0.45\linewidth]{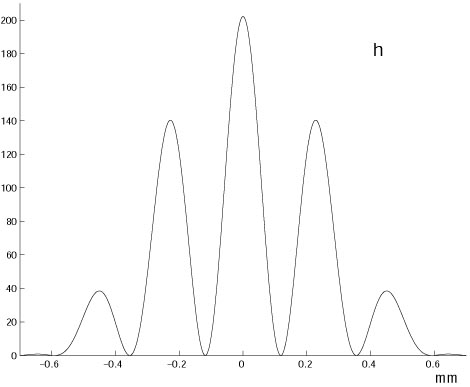}
\includegraphics[width=0.45\linewidth]{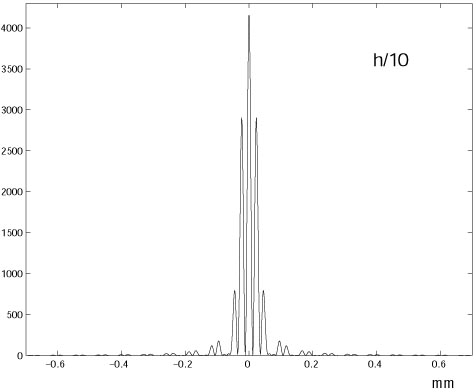}
\includegraphics[width=0.45\linewidth]{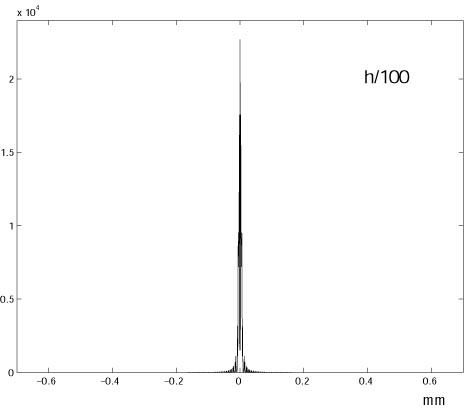}
\caption{\label{fig:convergenceMQ->MC-ddp-1}Cross section of the
probability density on the screen of detection when $h$ is divided
by 5, 10 and 100.}
\end{center}
\end{figure}

The figure~\ref{fig:convergenceMQ->MC-ddp-1} shows, when $h$ is
divided by 100, how the interference fringes are getting strongly
more and more narrow up to the distance between the slits.

\begin{figure}[h!]
\begin{center}
\includegraphics[width=0.45\linewidth]{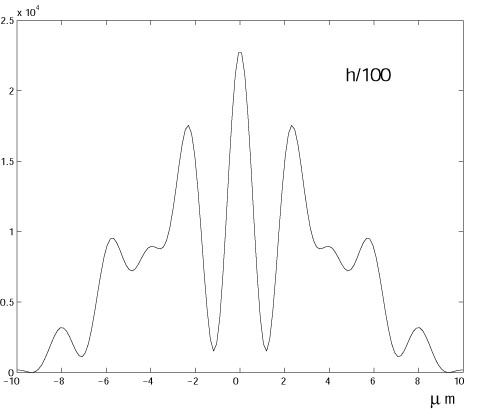}
\includegraphics[width=0.45\linewidth]{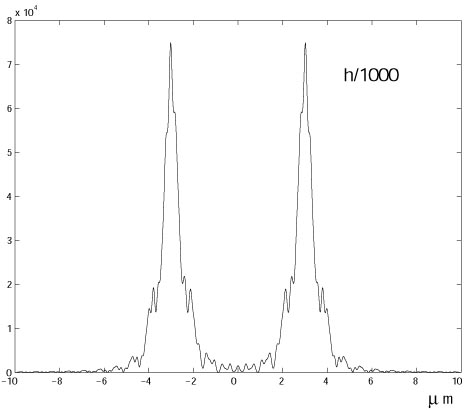}
\includegraphics[width=0.45\linewidth]{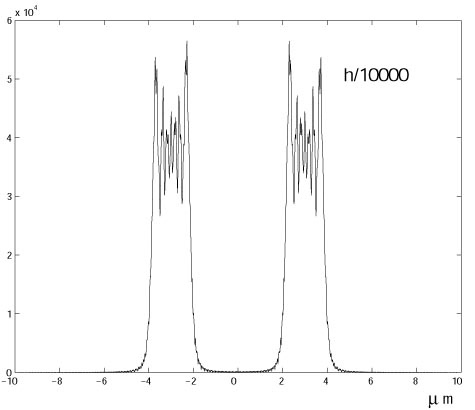}
\includegraphics[width=0.45\linewidth]{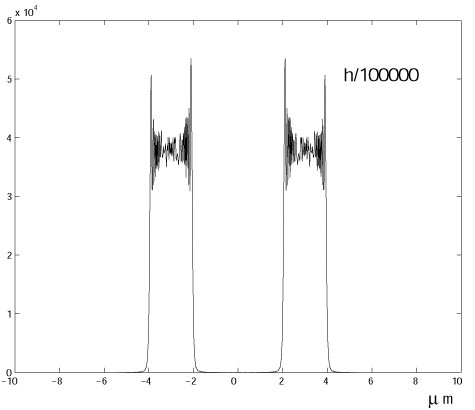}
\includegraphics[width=0.45\linewidth]{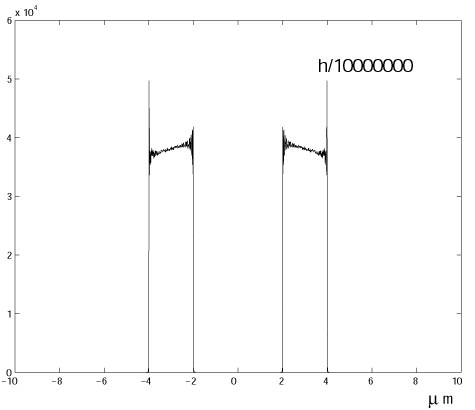}
\includegraphics[width=0.45\linewidth]{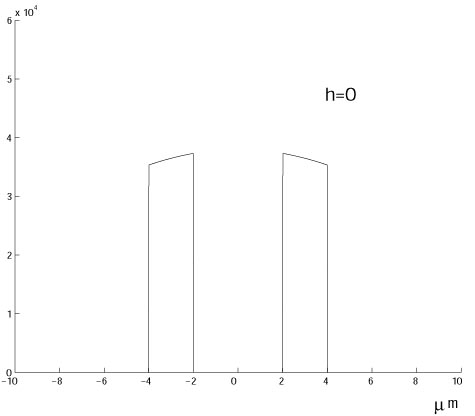}
\caption{\label{fig:convergenceMQ->MC-ddp-2}Cross section of the
probability density on the screen of detection when $h$ tends
towards zero.}
\end{center}
\end{figure}

The figure~\ref{fig:convergenceMQ->MC-ddp-2} shows, when $h$ is
divided by 1000, that the interference fringes disappear and the
probability density converges on the classical probability density
($h=0$).

\begin{figure}[h!]
\begin{center}
\includegraphics[width=0.45\linewidth]{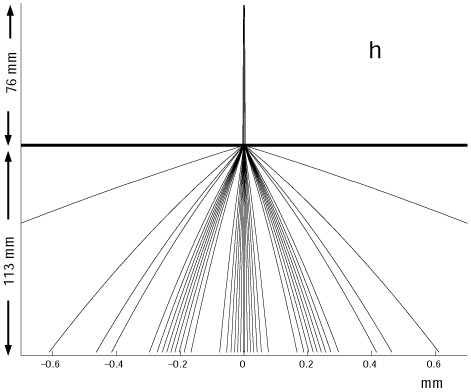}
\includegraphics[width=0.45\linewidth]{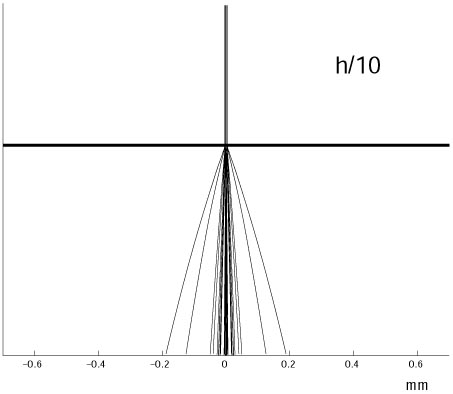}
\includegraphics[width=0.45\linewidth]{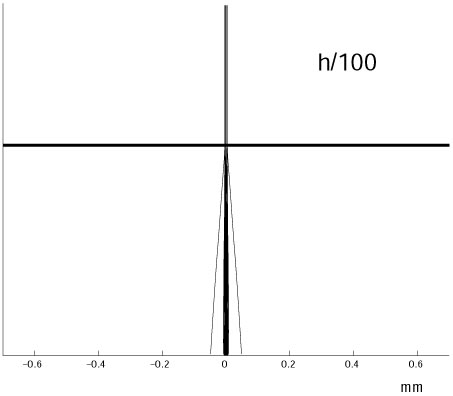}
\caption{\label{fig:convergenceMQ->MC-traj-1}Trajectories of atoms
for whole experience when $h$ is divided by 5, 10 and 100}
\end{center}
\end{figure}

The figure~\ref{fig:convergenceMQ->MC-traj-1} shows how the
trajectories become strongly narrower when $h$ is divided by 100.

\begin{figure}[h!]
\begin{center}
\includegraphics[width=0.45\linewidth]{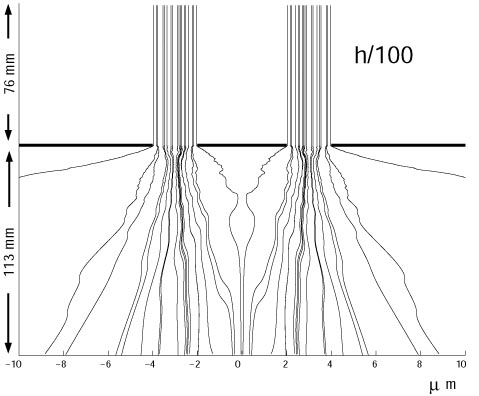}
\includegraphics[width=0.45\linewidth]{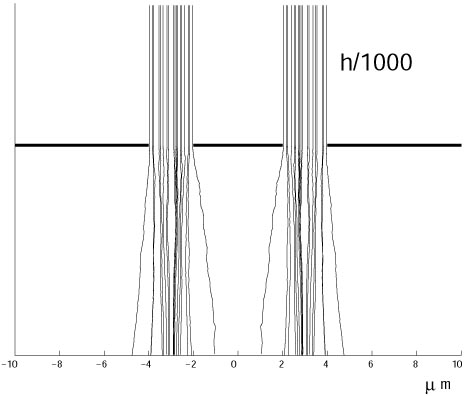}
\includegraphics[width=0.45\linewidth]{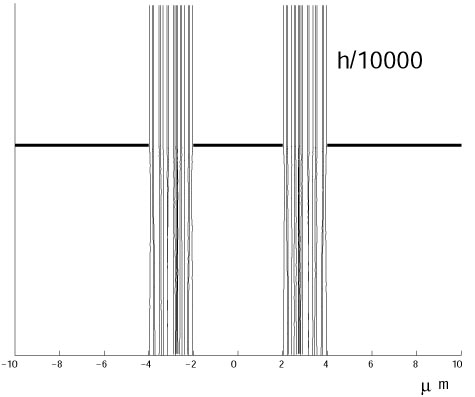}
\includegraphics[width=0.45\linewidth]{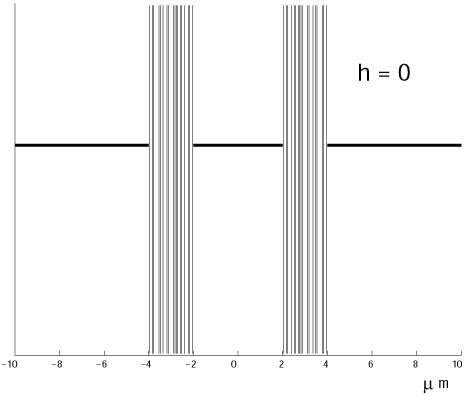}
\caption{\label{fig:convergenceMQ->MC-traj-2}Trajectories of atoms
for whole experience when $h$ tends towards zero.}
\end{center}
\end{figure}

The figure~\ref{fig:convergenceMQ->MC-traj-2} clearly shows the
convergence of the quantum trajectories to the classical
trajectories ($h=0$).

After showing results of the evolution of the probability density
and of the trajectories of atoms, it is difficult to throw down
the hypothesis of trajectory.

\textit{Everything seems to happen also as if atoms have well a
trajectory.}

\begin{acknowledgements}
We would like to thank the anonymous reviewers who provided
valuable critiques and constructive suggestions for better
presentation of the results.
\end{acknowledgements}

\end{document}